\documentclass[twocolumn,showpacs,preprintnumbers,amsmath,amssymb,aps,pre]{revtex4}

\usepackage{latexsym}
\usepackage{dcolumn}
\usepackage{supertabular,multirow}
\usepackage{epsfig}
\usepackage{bm}
\usepackage{subfigure}
\usepackage{amsmath}
\usepackage{graphicx}
\usepackage{color}

\definecolor{brown}{rgb}{0.59, 0.29, 0.0}
 \definecolor{orange}{RGB}{255,127,0}
\definecolor{brightube}{rgb}{0.82, 0.62, 0.91}

\newcommand{\bx}{{\bf x}}
\newcommand{\eps}{{\varepsilon}}
\newcommand{\half} {{\frac{1}{2}}}

\newcommand{\bU}{{\bf U}}
\newcommand{\bW}{{\bf W}}

\newcommand{\bF}{{\bf{F}}}
\newcommand{\bT}{{\bf{T}}}

\newcommand{\ba}{{\bf a}}
\newcommand{\bb}{{\bf b}}
\newcommand{\bc}{{\bf c}}
\newcommand{\bd}{{\bf d}}

\newcommand{\br}{{\bf r}}
\newcommand{\bv}{{\bf v}}
\newcommand{\bw}{{\bf w}}
\newcommand{\zhat}{{\bf \hat z}}
\newcommand{\yhat}{{\bf \hat y}}
\newcommand{\xhat}{{\bf \hat x}}
\newcommand{\rhat}{{\bf \hat r}}

\begin{document}

\title{Collective dynamics in a binary mixture of hydrodynamically coupled micro-rotors}

\author{Kyongmin Yeo$^{1,2}$, Enkeleida Lushi$^3$, Petia M. Vlahovska$^3$ }

\affiliation{$^1$IBM T.J. Watson Research Center, Yorktown Heights, NY 10598, USA\\
$^2$Division of Applied Mathematics, Brown University, RI 02912, USA\\
$^3$School of Engineering, Brown University, RI 02912, USA}


\begin{abstract}
We study numerically the collective dynamics of self-rotating non-aligning particles by considering a monolayer of spheres driven by constant clockwise or counterclockwise torques. We show that hydrodynamic interactions alter the emergence of large-scale dynamical patterns compared to those observed in dry systems. In dilute suspensions, the flow stirred by the rotors induces clustering of opposite-spin rotors, while at higher densities same-spin rotors  phase separate. Above a critical rotor density, dynamic hexagonal crystals form. Our findings underscore the importance of inclusion of the many-body, long-range hydrodynamic interactions in predicting the phase behavior of active particles. 
\end{abstract}


\maketitle

Systems of motile and interacting units can exhibit non-equilibrium phenomena such as self-organization and directed motion at large scales \cite{Marchetti13}. Theoretical studies of active matter report clustering \cite{Peruani06}, phase separations \cite{Fily12a, Redner13, Stenhammar13} and rotating structures \cite{Wensink14}. Some of these phenomena have been observed in experiments of bacterial suspensions \cite{Zhang10} or chemically-activated motile colloids \cite{Palacci10}. 

The collective motion of translating units such as bacteria has received much interest \cite{Marchetti13}. On the other hand, little is known about spinning units, partly because such systems were realized experimentally only recently. Active rotation of particles can be achieved using external forcing such as rotating magnetic fields \cite{Grzybowski00, Grzybowski02}, uniform electric fields (the Quincke rotation effect) \cite{Bricard13} or chemical reactions \cite{Wang}. Self-assembly from polymers by motile bacteria can create micro-rotors \cite{Schwarz12}. In biological systems, the dancing volvox \cite{Drescher09}, uniflagellar algae {\it C. reinhardtii} \cite{Kantsler14} and bacteria {\it T. majus} \cite{Petroff14} exhibit rotor-like behaviors. Rising interest in rotor systems generated theoretical studies exploring rotor pair dynamics \cite{Leoni10, Fily12b}, non-equilibrium structure formation \cite{Climent07}, dynamics at interfaces \cite{Llopis08, Lenz03},  rheology of suspensions \cite{Yeo10b, Jibuti12}, and phase separation driven by active rotation \cite{NguyenKEG14, Goto2015}.

Models of the collective  behavior  of active matter often neglect particle motion due to the flow stirred by the other particles \cite{Redner13, Ni13, Stenhammar13, NguyenKEG14}, tacitly assuming that the observed phase behavior of the ``dry'' system would persist in a system with fluid motion.
However in the viscosity-dominated world of colloidal-size particles, hydrodynamic interaction generates a long-range correlation, which can play an important role in the self-organization in many-body systems \cite{Campa, Baron, Tlusty}. 
For example, in the studies of micro-swimmers, it was found that the hydrodynamic interactions determine the collective motion of squirmers (self-propelled spheres with no aligning interaction) \cite{Zottl14} and the recently observed self-organization of bacteria into a macro-scale bidirectional vortex when confined inside a drop \cite{Wioland13} can only be explained by accounting for the fluid-mediated interactions \cite{Lushi14}. 
\begin{figure}
  \centering
  \includegraphics[width=0.48\textwidth]{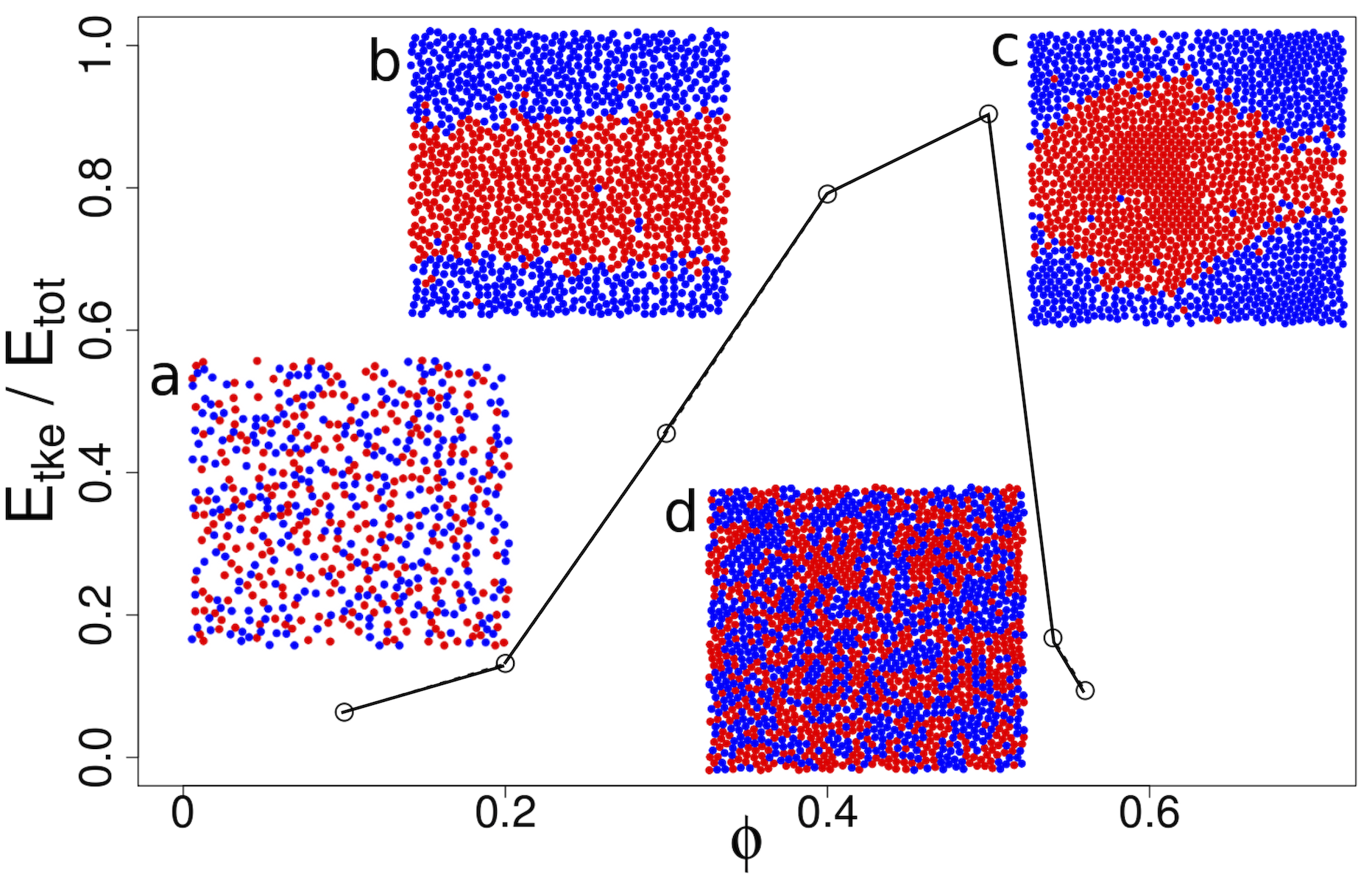}
    \vspace{-0.2in}
  \caption{ \footnotesize (Color online)  The ratio of the translational kinetic energy to the total kinetic energy $\kappa=E_{tke}/E_{tot}$ as a function of rotor density $\phi$. The insets are snapshots of simulations (50-50 mixture of clockwise(blue)--counter-clockwise(red) spinning rotors) with total density (a) $\phi$=0.20, (b) 0.40, (c) 0.50, and (d) 0.54. 
  Movies are available in the supplementary material \cite{supplem}.  }
  \label{fig1}
  \vspace{-0.2in}
\end{figure}
It is the hydrodynamic interactions that cause two point rotors spinning in the opposite direction to translate \cite{Fily12b} or undergo complex motions \cite{Lushi14b}, instead of remaining fixed in space \cite{NguyenKEG14}. While the importance of hydrodynamic interactions in micro-swimmers (linearly propelled units) has been appreciated, large and dense populations of rotors have not been studied and the robustness of observed phase behavior in the dry spinner system \cite{NguyenKEG14} remains an open question. 

{In this Letter, we  show that the hydrodynamic interactions between self-rotating non-aligning sphere particles have profound effects on self-organization.
 We consider monolayer suspensions of
 spherical rotors with clock- and counterclockwise spins suspended in liquid in a 3D domain \cite{supplem}. }
 At low densities, Figure \ref{fig1}a, a gas-like phase is observed with the rotors moving randomly in the stirred fluid. In contrast, in a dry system the spinners  remain fixed in place {(the frozen state in \cite{NguyenKEG14})}.
{As the particle density further increases, a phase-separated fluid state emerges (Figure \ref{fig1}bc) with large clusters of same-spin rotors manifesting as lanes or macroscopic vortical structures.  Past a critical particle density dynamic crystals composed of both types of rotors emerge, Figure \ref{fig1}d. }

{\em Particle motions.--}
 {We consider micro-rotors whose size is such that inertia is negligible (overdamped or Stokes flow regime), under the assumption of strong convection by the fluid flow. 
A rotor centered at  $\mathbf{x}_i$ with radius $a$ subjected to a torque {\bf T} generates a rotlet disturbance fluid flow $\mathbf{u}_R(\mathbf{x}, \mathbf{x}_i)= \mathbf{T} \times (\mathbf{x} - \mathbf{x}_i) a^3/|\mathbf{x} - \mathbf{x}_i|^3$ with velocity decaying slowly with the distance from the rotor as $\sim 1/r^2$.
The flow stirred by each rotor drags other rotors into motion. This is the essence of hydrodynamic interactions - a particle translates and rotates in response to the fluid flow generated by the motion of another particle. The rotors' positions and rotations evolve as \cite{supplem}
\begin{equation}
\label{eqM}
\begin{split}
\frac{d\mathbf{x}_i}{dt}&=  \sum_{j\neq i} \left[ \mathbf{u}_R (\mathbf{x}_i, \mathbf{x}_j) +O(\frac{a^7}{r^7}) \right]  + \sum_{j\neq i} \mathbf{F}^S_{ij} \\
\Omega_i &= \Omega_{0i} + \frac{1}{2} \nabla \times \mathbf{u}_R +O(\frac{a^8}{r^8}).
\end{split}
\end{equation}
$\mathbf{F}^S_{ij}$ are (purely repulsive) steric or excluded volume interactions between the particles. $\Omega_0=|\mathbf{T}|/8\pi \mu a^3$ is the rotation rate of an isolated rotor. Noise is neglected in Eq.\ref{eqM}, under the assumption of strong convection by the fluid flow $a^2\Omega_0/D_p \gg 1$ ($D_p$ is the particle diffusivity); for colloidal rotors of radius $1\, \mu m $ suspended in water this condition is met if  $\Omega_0>0.01 \, s^{-1}$, which is well below experimentally observed values \cite{Wang}. 

In dilute suspensions, where rotors are widely separated, the collectively-generated fluid flows are well-described by a superposition of the rotlet flows. However in dense suspensions the full hydrodynamic interactions and the inclusion of closer-range lubrication flows become complicated to resolve analytically and require the use of sophisticated numerical methods.}

The full hydrodynamic interactions between the rotors are computed using the force-coupling method. The long-range multi-body interactions are fully resolved by solving the Stokes equations with regularized low-order multipoles, while the short-range lubrication interactions are included from analytical solutions \cite{Yeo10a}.The force-coupling method has been successfully applied to study suspension flows \cite{Climent04,Yeo10f}. {For a description of the numerical method see the supplemental material \cite{supplem} and references therein.}

The numerical simulations of the {monolayer suspensions} are performed in a computational domain of $H_x \times H_y \times H_z = 80a \times 20a \times 80a$, in which $a$ is the particle radius and $y$ denotes the direction in which torques are applied. Periodic boundary conditions are used in the $x$ and $z$ directions. The particle monolayer is located at $y = 0$ and the computational box is bounded by rigid walls located at $y = \pm H_y/2$. The vertical separation is chosen big enough to guarantee that the wall boundary does not affect the monolayer dynamics. {Note that 
the rotors remain in the monolayer because the flow generated by their self-rotation does not induce particle  translation in a direction  normal to the monolayer  \cite{quasi2D}.}

We consider a 50:50 mixture of opposite-spin rotors with total volume fraction varying from $\phi = 0.1$ to $0.56$. For the monolayer suspension, the volume fraction is defined as $\phi = (\frac{4}{3}\pi a^3) N_p /  (H_x \times H_z \times 2a)$, in which $N_p$ is the number of the rotors
The number of the suspended rotors varies from $N_p = 306$ at $\phi = 0.1$ to $N_p = 1,712$ at $\phi = 0.56$. To model the active rotation, external torques are applied to the rotors in the $y$ direction. The magnitude of the external torque  is normalized by the fluid viscosity $\mu_0$ and the  reference angular velocity $\Omega_0$,  $T^* = T/8\pi\mu\Omega_0 a^3=\pm 1$. All of the simulations start from initial random configurations, generated by a molecular dynamics procedure. The dynamics are studied after the suspensions reach stationary states, typically about $t  \simeq O(10^4)$ from the initial random state (time is non-dimensionalized by $\Omega_0$).

\begin{figure}
  \centering
  \includegraphics[width=0.48\textwidth]{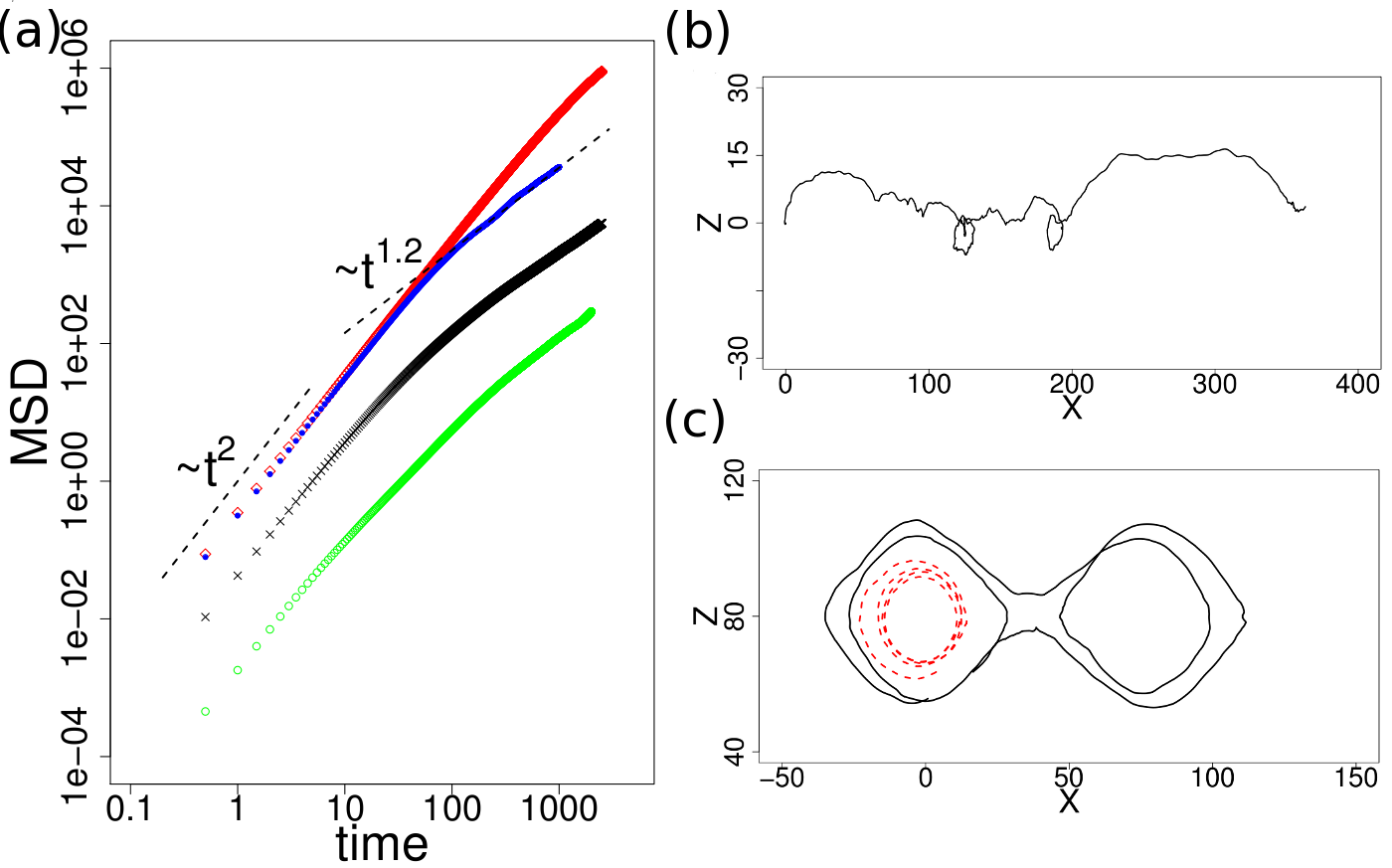}
  \caption{  \footnotesize (Color online)
  (a) MSDs for $\phi = $ 0.2 ($\times$), 0.4 ($\color{red}\diamond$), 0.5 ($\color{blue}\bullet$), 0.56 ($\color{green}\circ$). Representative trajectories for (b) $\phi  = 0.40$, (c) 0.50. The red and black lines refer to trajectories of different rotors. }
  \label{fig2}
  \vspace{-0.1in}
\end{figure}

{\em Hydrodynamic interactions and rotors translation.-- }  
Since dynamics here is overdamped and not noisy, the system behavior is controlled by only one parameter, the rotor density $\phi$.  
To assess the effect of the hydrodynamic interactions, we examine the conversion to translational kinetic energy ($E_{tke}$) of the rotational kinetic energy ($E_{rke}$) supplied to the suspension by the applied torque as rotor density increases. In the absence of hydrodynamic interactions, the rotors will remain fixed in space until random close packing is reached ($\phi_{rcp}\sim$0.56)  \cite{rcp}.  
{The flow generated by the rotating spheres moves them around and hinders their spinning \cite{supplem}. Accordingly, the translational kinetic energy is expected to increase with particle density}. However, Figure~\ref{fig1}  shows that the energy balance at steady state $\kappa = E_{tke} / E_{tot}$, where $E_{tot} = E_{rke} + E_{tke}$, depends non-monotonically on the  rotor density.

Initially, as the particle separation decreases with $\phi$, the  hydrodynamic interactions become stronger thereby increasing $\kappa$. 
{ In contrast to the dry 2D gear-rotor system in which $\kappa$ remains smaller than $2/3$ \cite{NguyenKEG14}, the equilibrium value set by equipartition, $\kappa$ here becomes larger than $2/3$ for $\phi  = 0.40$ and 0.50 as the suspensions phase-separate. In the phase-separated fluid regime, the suspensions develop large-scale collective motions, which contribute to the increase of $E_{tke}$.}
As the system approaches random close packing, $\kappa$ peaks at $\phi \simeq 0.50$ and decreases sharply afterwards, indicating a possible phase transition and change in the suspension microstructure. The $\kappa$-peak occurs prior to random close packing due to lubrication effects: the strong hydrodynamic resistance generated by the flow in the thin gap between particles effectively locks the rotors together leading to coherent motion.

{
The changes in suspension structure are also suggested by the behavior of the mean-squared rotors displacement $MSD=\langle |\bx_i(t)-\bx_i(0)|^2\rangle/a^2$, shown in Figure \ref{fig2}a. Hydrodynamic interactions give rise to random rotor motion, which in the short-time limit exhibits the typical ballistic $\sim t^2$ behavior. 
However, at intermediate times $100 < t < 1000$, MSD changes from diffusion $\sim t$ at $\phi = 0.2$ to superdiffusion at $\phi = 0.4$ and 0.5.  
{At short timescale, MSDs for $\phi = 0.4$ and 0.5 are almost the same, as both systems are in the phase-separated fluid states. In the long-time limit, however, MSD depends on the large-scale collective motion.
For $t > 50$, MSD for $\phi = 0.4$ grows at a much faster rate than $\phi = 0.5$.}
The superdiffusivity is due to  L\'{e}vy flights of the rotors \cite{Solomon93} seen in  Figure  \ref{fig2}bc. The trajectories show that  at $\phi = 0.4$ individual rotors move longer distances and circulate less in the macroscopic vortices compared to $\phi = 0.5$.
At $\phi = 0.56$ the MSD is dramatically reduced due to crystal formation.  Unlike to the dry gear-like rotor system \cite{NguyenKEG14}, caging is not obvious in the MSD.
}

{\em Spin Segregation.--} 
In the range of densities below the sharp drop in $\kappa$ (i.e., $\phi \le 0.5$), the rotors form dynamic assemblies \cite{supplem} which in Figure \ref{fig1}bc are indicated as ``phase-separated fluid''.
To quantify this tendency to cooperative motion we compute the number densities of the opposite-spin and same-spin rotors within distance $r$
\[
\lambda^\pm(r) = \left\langle \frac{1}{N} \sum_{i=1}^N \left\{ \frac{ \sum_{j=1, j \neq i}^{N} H(r - |\bm{d}_{ij}|)\delta(T_i \pm T_j)}{n \pi r^2 (2a)}   \right\} \right\rangle.
\]
 $H(x)$ is the Heaviside function, $\delta(x)$ is the Dirac measure, $N$ is the number of the suspended rotors, $|\bm{d}_{ij}|$ is the distance between the  $i-th$ and $j-th$ rotors, and $n$ is the number density. 
{$\lambda^\pm$ 
are related to the pair distribution functions, $g_{AA}(r)$ and $g_{AB}(r)$ as $\lambda^- (r) \sim \int \left(g_{AA}\right) r dr$ and  $\lambda^+(r) \sim \int \left(g_{AB}\right) r dr$; it can be interpreted as the average number of coherently moving neighbors \cite{Wysocki14}.}

\begin{figure}
  \centering
  \includegraphics[width=0.48\textwidth]{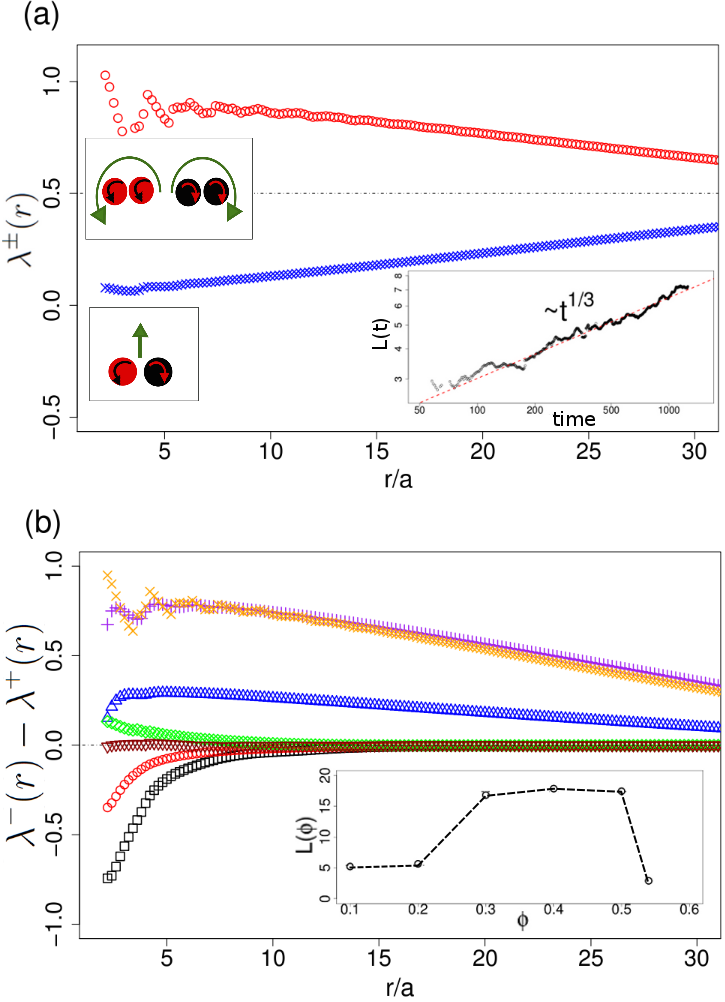}
    \vspace{-0.1in}
  \caption{\footnotesize (Color online) (a) Normalized partial number density of  same-spin ($\lambda^-;~{\color{red}\circ}$) and opposite-spin ($\lambda^+;~{\color{blue}\times}$) rotors for $\phi = 0.5$ at steady state. Insets illustrate co- and counter-rotating particles and their joint rotation or translation. The inset shows a $\sim t^{1/3}$ growth of the lengthscale  from the initial random configuration.
(b) Average density difference between coherently moving same-spin and opposite-spin rotors, $\lambda^--\lambda^+)$, for $\phi = 0.1$ ($\square$), 0.2 ({\color{red}$\circ$}), 0.3 ({\color{blue}$\triangle$}), 0.4 ({\color{brightube}$+$}), 0.5 ({\color{orange}$\times$}), 0.54 ({\color{green}$\Diamond$}), and 0.56 ({\color{brown}$\nabla$}). The inset shows the final integrated lengthscale $L$ as a function of $\phi$.  }
   \label{fig3}
    \vspace{-0.1in}
\end{figure}

Figure \ref{fig3}a illustrates $\lambda^\pm$ for $\phi = 0.5$. $\lambda^-(r)$  exceeds $\lambda^+(r)$ at small separations $r$ implying clustering of the same-spin rotors. In the far field ($r > 30a$), eventually the number densities of the same- and opposite-spin rotors become the same. The average cluster size can be characterized by the lengthscale over which the correlations between the rotors die out, $L(t) = \int r \left(\lambda^--\lambda^+\right) \, dr / \int \left(\lambda^--\lambda^+\right)  \, dr$. 
Figure \ref{fig3}a shows that $L$ grows as $\sim t^{1/3}$, which eventually saturates to the value shown in the inset of Figure \ref{fig3}b (17.4 in this case). 
The exponent of $1/3$ is surprising as it is usually associated with coarsening dynamics in the absence of hydrodynamics. Hydrodynamic interactions are however known to give rise to diffusive behavior in suspension flows \cite{Sierou04,Yeo10c, Davis}.

A more careful examination of the clustering shows that, in dilute suspensions ($\phi \le 0.2$), rotors of opposite-spin tend to pair-up. 
 Figure \ref{fig3}b shows that the difference between $\lambda^-$ and $\lambda^+$ reverses sign, indicating clustering of same-spin rotors, as the density increases above $\phi \sim 0.2$.  The change of microstructure occurs because while at low densities the separation between rotors is large thereby allowing rotors to explore more space by translation (a pair of opposite spin rotors translates \cite{supplem}), at higher densities assemblies that are less obstructing to the motion of other rotors are preferred (a pair of same spin rotors orbits around each other \cite{supplem}).
At $\phi = 0.4$ and 0.5, where complete separation occurs, $\lambda^- -\lambda^+$ are almost identical for $r > 5a$. As $\phi$ increases further, $\phi \ge 0.54$, $\lambda^- - \lambda^+$ becomes close to zero, suggesting there is no or very weak preferential aggregation of the rotors. In the inset, $L$ is shown as a function of $\phi$. Spin segregation is captured by the integrated length-scale which increases sharply at  $\phi=0.2$ and  drops rapidly for $\phi \ge 0.54$.\\

\begin{figure}
  \centering
  \includegraphics[width=0.48\textwidth]{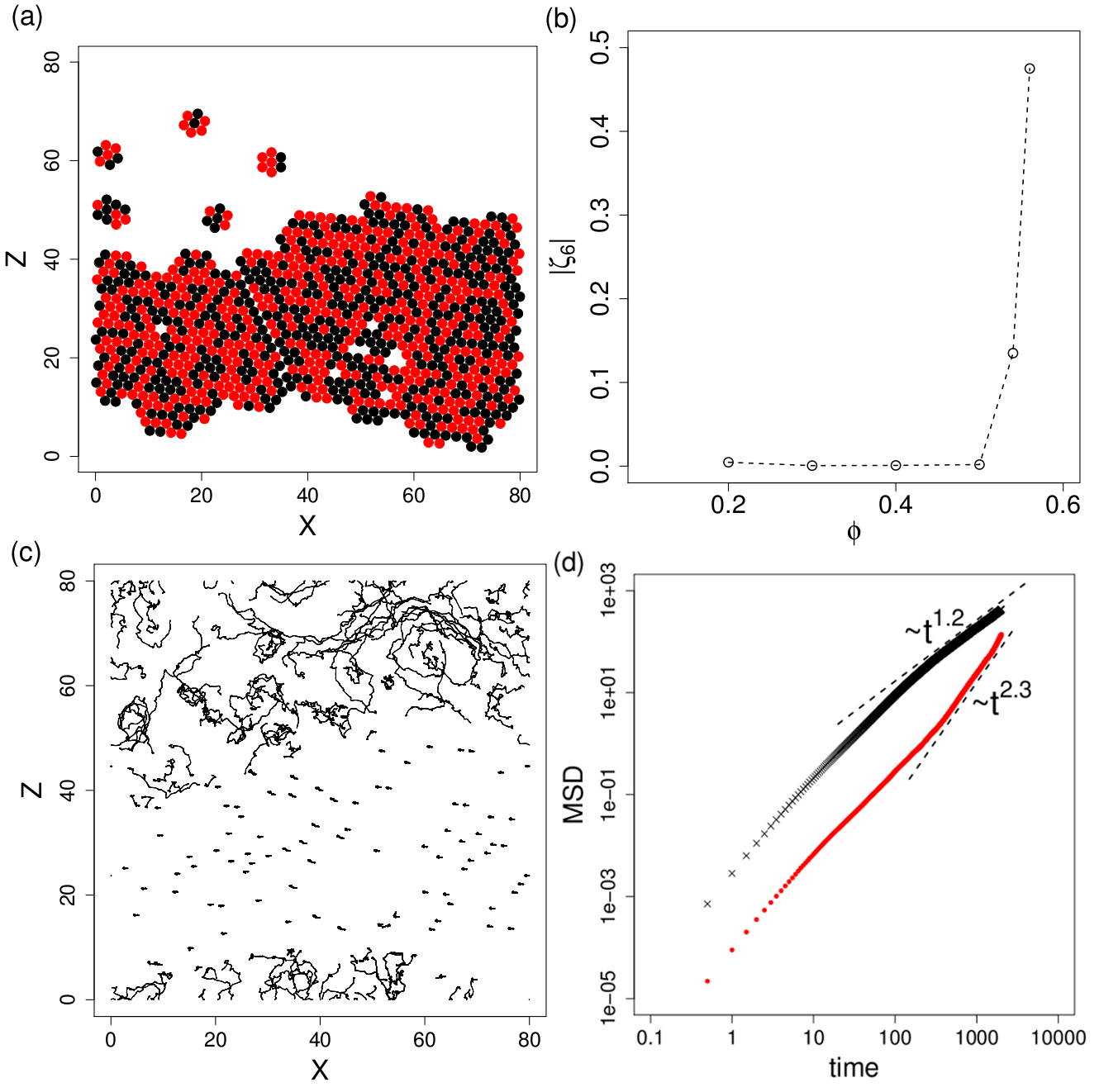}
  \vspace{-0.2in}
  \caption{\footnotesize(Color online) (a) Crystal structures formed in a suspension at $\phi$ = 0.56. The black and red circles denote the rotors rotating clockwise and counter-clockwise respectively. Note that for clarity only the rotors in crystals are shown. (b) Hexagonal bond-orientational order parameter $\zeta_6$ as a function of $\phi$. (c) Sample trajectories for $t = 0 \sim 600$ for $\phi = 0.56$. (d) MSDs of the rotors initially in crystal (${\color{red}\bullet}$) and in fluid regions ($\times$). }
  \label{fig4}
  \vspace{-0.1in}
\end{figure}

{\em Crystals.--} At high density, rotors form crystals of hexagonal symmetry, see Figure \ref{fig4}.a. {The crystals are composed of rotors of either spin, and no spin segregation is observed for the duration of the simulations.}
The fraction of the crystal phase increases with density, and at $\phi = 0.56$ the crystal structure occupies roughly half of the computational domain.
The formation of the crystals is tracked by an order parameter, $0\le \zeta_6 \le1 $, which measures the average sixfold bond orientational order of the rotors;
\[
\textstyle{\zeta_6 = \left\langle \frac{1}{N} \sum_i^N \left( \frac{1}{N_b} \sum_j^{N_b} e^{6 \theta_{ij} \mathrm{i}}  \right)  \right\rangle.}
\]
$\theta_{ij}$ is the azimuthal angle of $\bm{d}_{ij}$ and $N_b$ is the number of the neighboring rotors ($|\bm{d}_{ij}| < 2.05a$).  $\zeta_6$ is zero for an isotropic system and one for a perfect hexagonal crystal. Figure \ref{fig4}b shows that $\zeta_6$ is almost zero up to $\phi = 0.5$ and increases rapidly from $\phi \simeq 0.54$, which corresponds where the sudden drop of $\kappa$ occurs (see Figure \ref{fig1}).

Interestingly, even in the presence of crystals the rotors exhibit superdiffusive behavior with an exponent $\sim t^{1.2}$ { at the intermediate timescale}, see Figure \ref{fig2}a.
Figure \ref{fig4}c illustrates trajectories of 170 randomly selected rotors. Particle mobility is much lower in the crystal region than in the fluid region: even though the trajectories are shown for a relatively long period $t = 0 \sim 600$, rotors located in a crystal move only very short distance, usually less than a particle diameter, while rotors in the fluid region travel considerably longer distance ($>15 a$). The difference in mobilities is also evident from Figure \ref{fig4}d, which compares the MSDs for the rotors initially in a crystal and in a fluid region.
MSD for the rotors in a crystal grows very rapidly $\sim t^{2.3}$ for $t > 400$. {The rapid growth of MSD seems related to structural re-arrangements, \emph{i.e.}, large-scale motions of crystal and escape of the rotors in the crystal to the fluid region. Note that the crystal structure dynamically melts, re-assembles, and moves, see movies in \cite{supplem}.}

{\em Conclusions and outlook.--} 
Suspensions of active particles exhibit complex phase behavior \cite{Marchetti13} 
and self-translating particles have attracted extensive studies \cite{Wysocki14, Redner13, Fily12a, Zottl14}.
Here we show that self-rotating particles are driven by hydrodynamic interactions into mobile clusters and crystals even in the absence of self-propulsion or ambient flows. The resulting collective dynamics is very different from that observed in a dry system \cite{NguyenKEG14}.

The effect of the hydrodynamic interactions is assessed by observing the conversion rate of the rotational to the translational kinetic energies ($\kappa$).
$\kappa$ initially increases with the rotor density, and eventually exhibits a sudden drop at $\phi \simeq 0.54$. For $\phi \ge 0.54$, the active rotors start to form crystal structures, which are responsible for the sudden drop of $\kappa$.  In contrast, the dry, no-noise system of gear-like rotors \cite{NguyenKEG14} exhibits $\kappa=0$ (in the frozen state) followed by monotonic increase of $\kappa$ above a critical density $\phi_c$ corresponding to about $0.5$ in our notation.We found that the opposite-spin rotors tend to stay close at low $\phi$, whereas for $\phi > 0.2$ separation into fluid phases of same-spin rotors occurs.
 All of these suggest that multi-body hydrodynamic interactions play a significant role in the collective dynamics and phase behavior of suspensions of active rotors and these effects should not be neglected in studies of similar active systems. For example, hydrodynamic interactions could influence or drive the formation of the peculiar dynamical structures experimentally observed at the interface of drops covered with colloidal particles \cite{Dommersnes13, Ouriemi14}. 

In this Letter we considered only torques that are perpendicular to the particle monolayer. 
Due to the symmetry of the generated flows, the particles remain confined to the monolayer and do not move transversely. 
In experimental systems, for example Quincke rotors \cite{Salipante13, Das13}, it is not the case that torques stay in one direction or even constant, as the particle rotation is dependent on the full electro-hydrodynamics. Although restricting the rotational motion to one direction in experiments is challenging, it is not impossible and our study suggests potentially intriguing experiments. Another problem that remains relatively unexplored is that of using spinners and rotors for transport and mixture of passive particles \cite{NguyenKEG14}. Finally, this work with rotor-monopoles serves as a solid basis to treat rotor-dipoles, which are commonly encountered in biology, e.g., swimming bacteria with rotating flagella or a cytoskeletal torque dipole consisting of two actin filaments and myosin motors \cite{Furthauer13}.

\newpage
{\em Acknowledgements:} EL and PV acknowledge support from the NSF through CBET award 1437545.

\appendix
\newpage

\begin{center}
{\large \textbf{Supplementary Material}}\\
{\bf{Collective dynamics in a binary mixture of hydrodynamically coupled micro-rotors}}\\
K. Yeo, E. Lushi, P. M. Vlahovska
\end{center}

\begin{figure*}
\centerline{\includegraphics[width=0.88\textwidth]{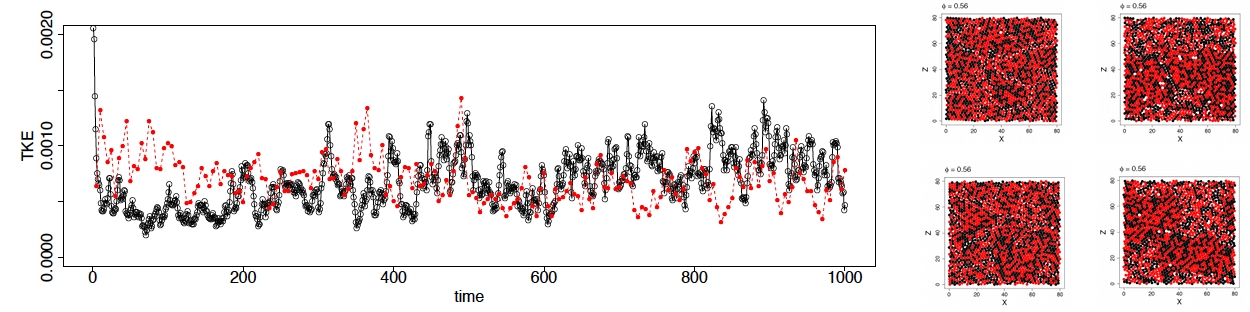}}
\vspace{-0.2in}
\caption{ At $\phi=0.56$ a well mixed state of rotors with randomly distributed spins remains well-mixed. The figure shows the TKE (translational kinetic energy) history from two different initial configurations and confirms that in this case no spin segregation occurs and that the system is equilibrated. }
\vspace{-0.2in}
\label{fig1ss}
\end{figure*}

\section{Movies}

Movies phi20, phi40, phi50, and phi56 illustrate the long-time suspension dynamics corresponding to Figure 5a-c, and Figure 4.a. Movie phi50\_transient shows the phase separation and the growth of the  cluster scale $L$ from the random initial configuration $0<t<2740$.  The duration of the movies is 1000 for phi20, phi40, and  phi50, and 3500 for phi56

\section{Three-dimensional nature of the monolayer dynamics }

\begin{figure*}
\centerline{\includegraphics[width=0.6\textwidth]{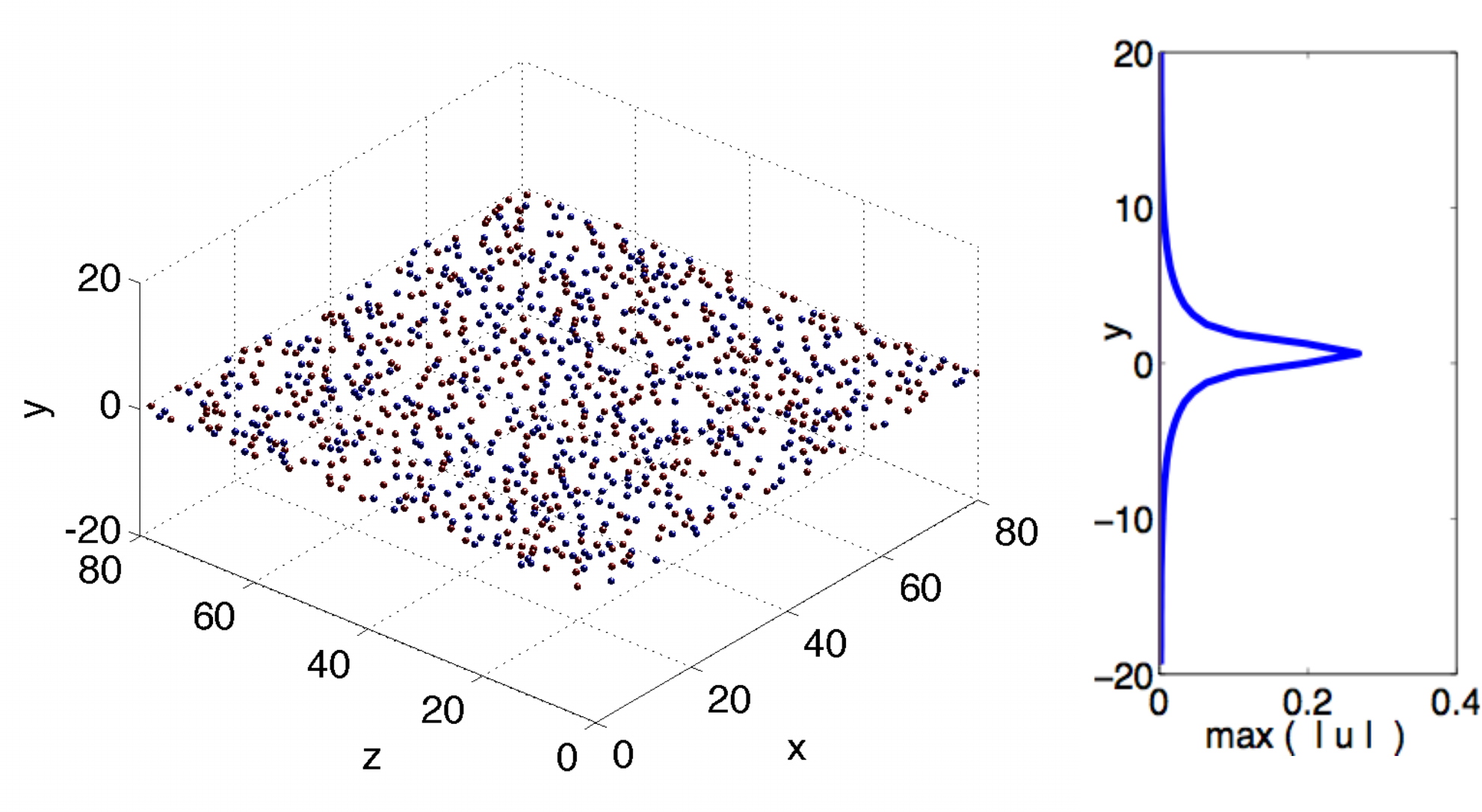}}
\vspace{-0.2in}
\caption{ (a) 3D view of the monolayer and the computational domain. (b)Magnitude of the fluid velocity along the direction normal to the monolayer. The velocity decays  to zero at distance about 10 rotor radii from the monolayer, which shows  that the flow generated by the spinning rotors is long-ranged. }
\label{fig2ss}
\end{figure*}

The monolayer is embedded in a 3D fluid. The flows in the surrounding fluid (above and below the monolayer) are unobstructed and hence long ranged, as illustrated in Fig. \ref{fig2ss}. The hydrodynamic interactions mediated by these flows engender three-dimensionally in our system, which is masked by the fact that the spheres remain in the monolayer. 

The spheres do not leave the monolayer since the flow generated by their self-rotation does not induce sphere translation in a direction perpendicular to the monolayer. In contrast to \cite{Zottl14}, we do not need walls to confine the particle motion to a plane.   Accordingly, characterizing our system as quasi-2D may be misleading.

\section{Hydrodynamic interactions of two rotors}

A particle in the fluid moves in response to a force acting on it (e.g., buoyancy)  and  the hydrodynamic drag due to the flow created by the 
other moving particles. Because  the flow about a microparticle is in the Stokes (overdamped) regime,  the translational velocity and rotation rate of the particles   are linearly related to the forces and torques exerted by the fluid on the particles via the mobility matrix \cite{KimKarrilla, Schmitz} (note that the force/torque exerted by the fluid on the particle has the opposite sign of the force/torque exerted by the particle on the fluid)
\begin{equation}
\left( \begin{array} {c}
  \bv^\infty- \bU_p\\
\bw^\infty - \bW_p
 \end{array} \right ) 
=
\left( \begin{array} {ccc}
   \ba& \tilde \bb\\
   \bb& \bc& 
 \end{array} \right ) 
 \left( \begin{array} {c}
 \mu^{-1} \bF\\
 \mu^{-1}\bT
 \end{array} \right ) 
\end{equation}
 $\bv^\infty$ and $\bw^\infty$ are the velocity and rotation rate of applied flow, and $\mu$ is the viscosity of the suspending fluid.

It is instructive to precede the detailed computational treatment of many rotors
with an  outline of the pair-wise hydrodynamic interactions, which would provide some intuition about the hydrodynamic effects.

In the case of two  force--free (e.g., neutrally--buoyant) particles each subjected to a constant  torque, $\bT^n$ ($n=1,2$), in the absence of background flow (initially quiescent fluid), $\bv^\infty=\bw^\infty=0$ the translational and angular velocities  of particle$n$ are
\begin{align}
-\mu \bU^n&=\tilde \bb^{n1}\cdot \bT^1+\tilde \bb^{n2}\cdot \bT^2\\
-\mu \bW^n&= \bc^{n1}\cdot \bT^1+ \bc^{n2}\cdot \bT^2
\end{align}

Following Kim and Karrila's notation, let us use $\alpha, \beta$ to denote the particles and $i,j,k$ to denote the $x,y,z$ components. 
A unit vector $\bd$ points along the axis connecting the particles,
\begin{align}
\bd=\rhat_{12}
\end{align}
and the center-to-center distance is $r$.

The mobility tensors are  listed in \cite{KimKarrilla}. 
From symmetry $\tilde b^{(\beta\alpha)}_{ji}=b_{ij}^{(\alpha \beta)}$. 
\begin{align}
b_{ij}^{(\alpha \beta)}&=y^b_{\alpha \beta}\epsilon_{ijk}d_k \\
c_{ij}^{(\alpha \beta)}&=x^c_{\alpha \beta} d_i d_j+y^c_{\alpha \beta}\left(\delta_{ij}- d_i d_j\right)
\end{align}

\subsection{Rotor translation}
The translational velocities of the rotors $n=1,2$ are
\begin{align}
-\mu \bU^n=y^b_{1n}\bd \times  \bT^1+y^b_{2n}\bd \times \bT^2\,.
\end{align}
Thus, the relative velocity $\dot \br_{12}=\bU=\bU^1-\bU^2$ is
\begin{align}
\dot\br_{12}= \bU=-\mu^{-1}\left(y^b_{11}-y^b_{12}\right) \rhat_{12} \times \left( \bT^1+ \bT^2\right)\,.
\end{align}
The rotors separation remains the same if the center-to-center axis is perpendicular to the axis of rotation.

\begin{figure}
\vspace{-0.in}
\centering
\includegraphics[width=0.45\textwidth]{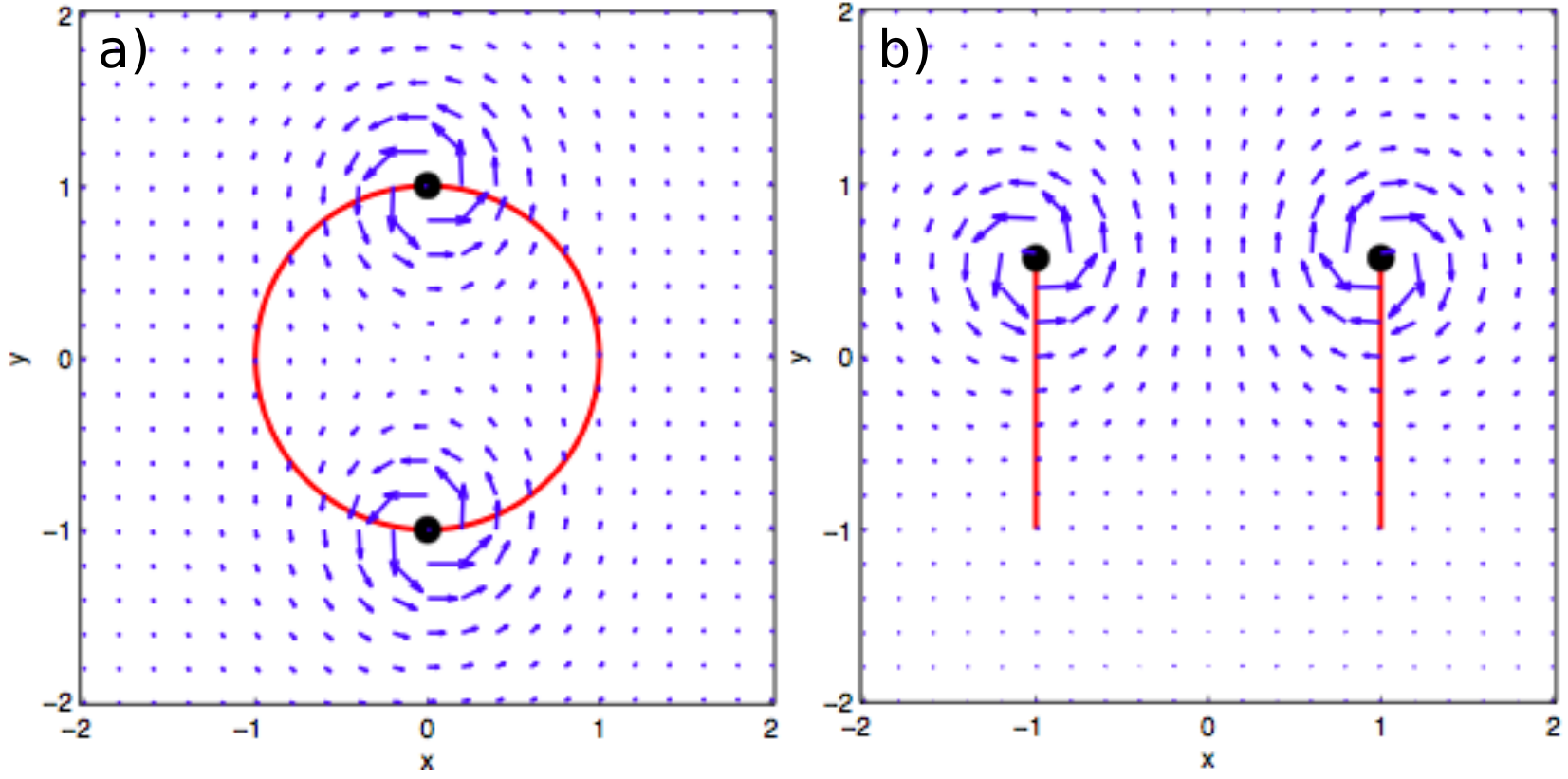}
\vspace{-0.1in}
\caption[width = 0.55\textwidth]{  (Color online) (a) A pair of same-spin rotors circulates around their center of mass. (b) A pair of opposite-spin rotors self-propels in the direction perpendicular to their separation. The blue arrows show the fluid flow, the red lines show the rotors trajectories. Figure taken with permission from \cite{Lushi14b}.}
\vspace{-0.in}
\label{fig3ss}
\end{figure}

The motion of the center of the mass $\dot\br_{cm}=\half(\bU_1+\bU_2)$ is given by
\begin{align}
\dot \br_{cm}= \bU_{cm}=-\mu^{-1}\half \left(y^b_{11}-y^b_{12}\right) \rhat_{12} \times \left( \bT^1- \bT^2\right)\,.
\end{align}
Let us consider spherical rotors spinning around the x-axis, $\bT^1=\eps^{-1} \bT^2=-\tau \yhat $
\begin{align}
\dot \br_{cm}=\half A(r)(1-\eps)\rhat_{12} \times \yhat\quad \dot \br_{12}=A(r)(1+\eps)\rhat_{12} \times \yhat
\end{align}
where 
\begin{align}
A(r)=\frac{\tau}{8\pi a^2\mu}\left( \frac{a}{r}\right)^2\left(1-\frac{13}{2}\left(\frac{a}{r}\right)^5+...\right)
\end{align}
Hence, if the rotors are of opposite spin $\eps=-1$ they undergo net translational motion while their separation remains constant. In the case of co-rotating rotors $\eps=+1$ the two rotors orbit each other (see Fig. \ref{fig3ss}).

We have utilized the expressions from \cite{KimKarrilla, Schmitz}:
\begin{align}
y_{11}^b &=\frac{1}{4\pi a^2}\left( f_7 \left(\frac{a}{2r}\right)^7+..\right)\\
 y_{12}^b &=\frac{1}{4\pi a^2}\left( f_0+f_2 \left(\frac{a}{2r}\right)^2+..\right)
\end{align}
where $f_0=0\,, f_2=-2$ and $f_7=160+48=208$.

\subsection{Modification in rotation rate}
An isolated rotor subject to torque $\bT$ rotates with rotation rate $\Omega_0=T/(8\pi \mu a^3)$. The flow produced by other rotors modifies the rotor angular velocity.

Considering the two rotors, $n=1,2$,  spinning around the x-axis due to applied torques $\bT^1=\eps^{-1} \bT^2=-\tau \yhat $, the angular velocity of rotor $n$ is:
\begin{align}
 \bW^n=\tau\mu^{-1}(\bc^{n1}+\bc^{n2} \eps) \cdot\yhat\,, \quad \bW=\Omega_x\xhat+\Omega_y\yhat+\Omega_z\zhat
\end{align}

\begin{align}
\Omega^1_i=\frac{\tau}{\mu}\left[\left (x_{11}^c+\eps x_{12}^c\right)d_id_j\hat x_j+\left(y_{11}^c+\eps y_{12}^c\right)\left(\delta_{ij}-d_id_j\right)\hat x_j\right]
\end{align}
since $\hat x_2=1$, $\hat x_1=\hat x_3=0$ and $d_3=1$, $d_1=d_2=0$ we find that
\begin{align}
\label{eqW}
\Omega^1_y &=\frac{\tau}{\mu}\left[\left (x_{11}^c+\eps x_{12}^c\right)\right] \nonumber \\
&=\frac{\tau}{8\pi\mu a^3}\left (1-\frac{\eps}{2} \left(\frac{a}{r}\right)^3 \right)+O\left(\left(\frac{a}{r}\right)^8\right)
\end{align}
where we have taken the values for $x^c$ from \cite{KimKarrilla, Schmitz}. Note that same result (Eq. \ref{eqW}) can be obtained from Faxen's law as given by Eq. 1 in the main text. 

If the rotors are same spin, $\eps=+1$, the hydrodynamic interaction retards their rotation. 

\subsection{Rotation hindrance in a binary mixture of rotors (50\% with $\eps=1$ and 50\% with $\eps=-1$)}

\begin{figure}
\centerline{\includegraphics[width = 0.24\textwidth]{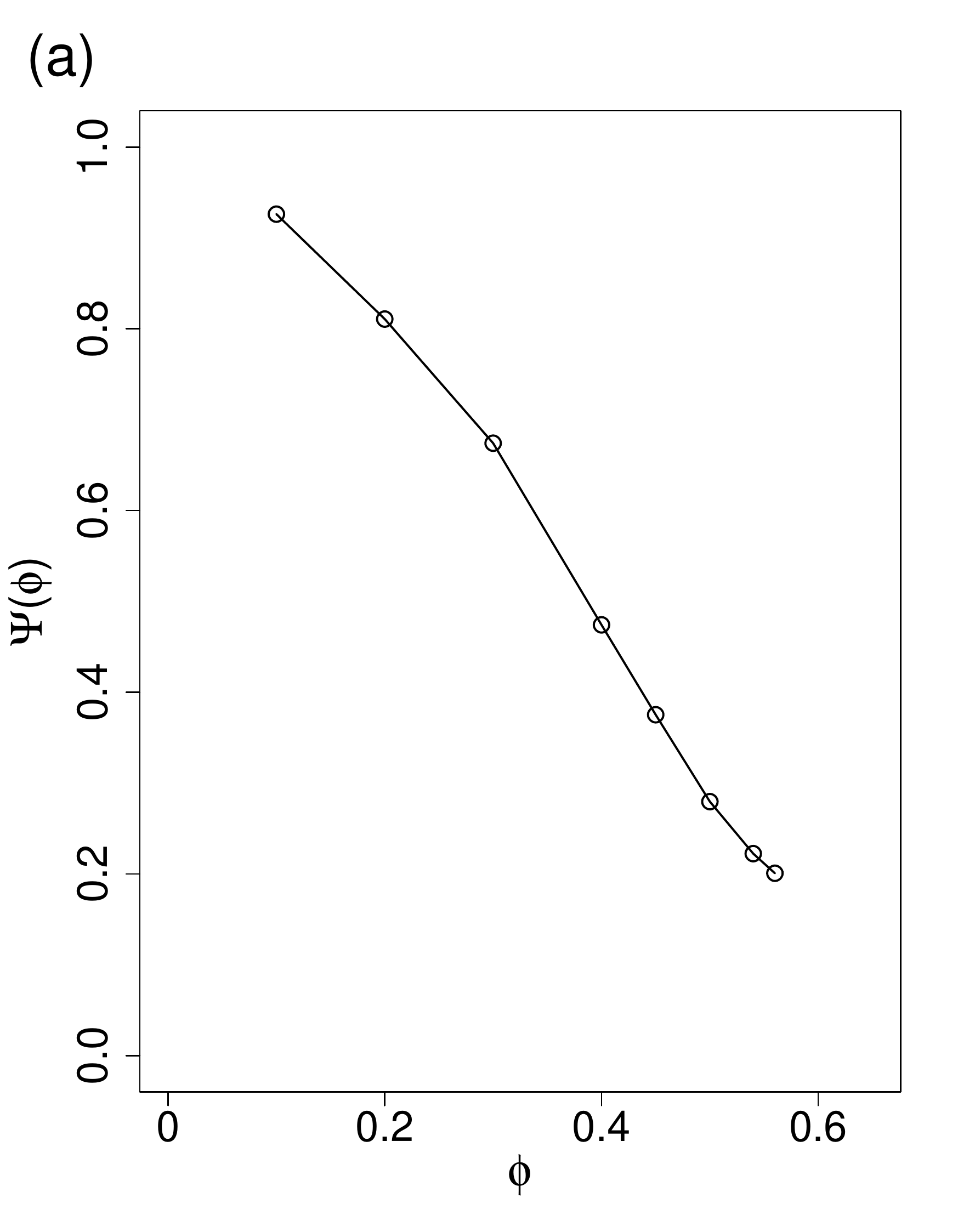}}
\vspace{-0.2in}
\caption{ The angular velocity hindrance function $\Psi$ as a function of the rotor density $\phi$.  }
\label{fig4ss}
\end{figure}

The angular velocity of a rotor  decreases with particle volume fraction  \cite{Brenner}. To quantify the hydrodynamic resistance to the intrinsic torque, the angular velocity hindrance function is defined as $\Psi = \langle |\Omega_y| \rangle / \Omega_0$, in which $\langle \cdot \rangle$ denote an ensemble average. Note that the applied torques are in the $y$ direction, as considered in the main text. Figure \ref{fig2}  shows that the angular velocity hindrance function ($\Psi$) monotonically decrease with the volume fraction $\phi$. 

\subsection{Effect of rotor shape}

\begin{figure*}
  \centering
  \includegraphics[width=0.88\textwidth]{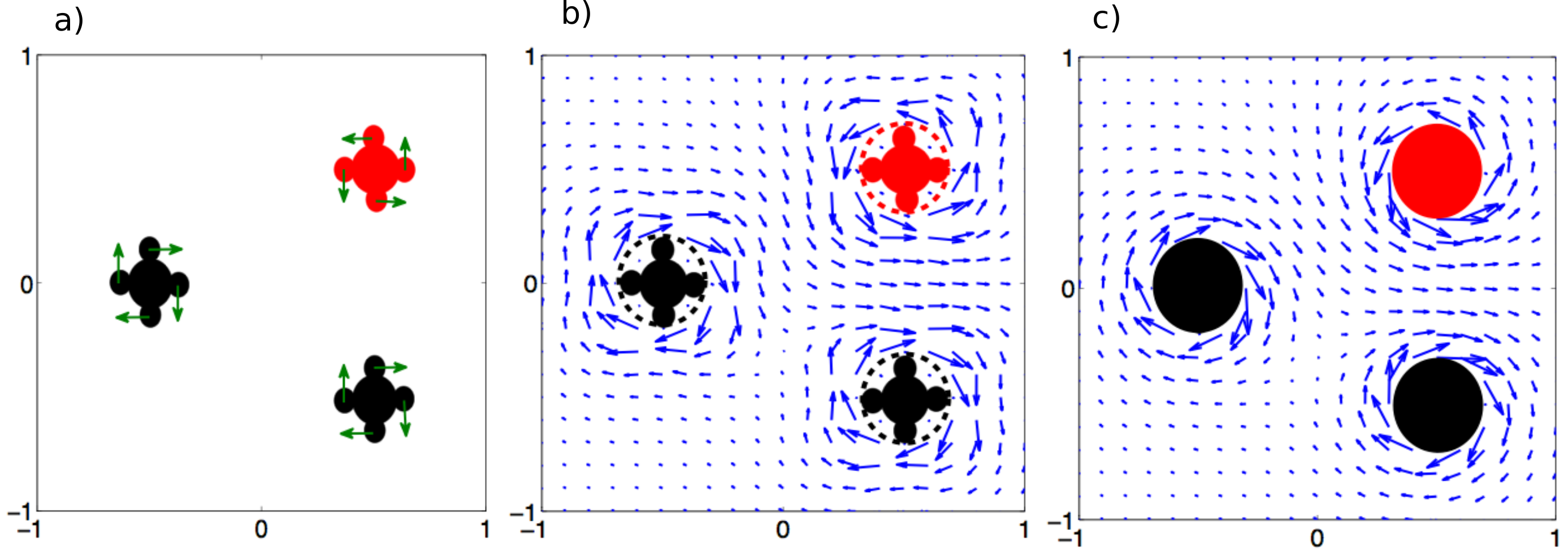}
  \vspace{-0.2in}
  \caption{ (a) Three gear-like spinning particles as considered by Nguyen {\it et.al.} \cite{NguyenKEG14} with arrows at the gear-ears showing rotation motion. (b)  3D fluid flow generated by the  gear-like particles, obtained as a superposition of Stokeslets applied at the gear-ears. (c) Fluid flow generated by three spherical rotating particles obtained by a superposition of three rotlets at the sphere centers. The sphere size is the effective size of the gear-like particles shown in (b).}
  \label{fig5ss}
\end{figure*}

In this paper we concentrate on spinning spheres  because we are interested in the collective dynamics of non-aligning particles. 
However, our results also apply to dilute and semi-dilute suspensions of more complex shaped rotors because the  flow due to any rotating particle looks like rotlet flow in the far field. Figure \ref{fig5ss} compares the  flow field around spinning spheres and  spinning four-gears particles such as those considered by Nguyen {\it et al.} \cite{NguyenKEG14}.
 The fluid flows generated by the gear-like spinners (shown in Figure \ref{fig5ss}.b) can be approximated by a superposition of the flows due to Stokeslets located at the gear-ears  as shown in  Figure \ref{fig5ss}.a.  The fluid flow generated by three rotating spheres with radii equal to the effective radii of the gear-like spinners can be computed by a superposition of three rotlets applied at the sphere centers, and is shown in Figure \ref{fig5ss}.c. The fluid flow fields in Figures \ref{fig3}.b and c look remarkably similar already at one radius away from the particles, thus the leading order and far-field fluid flows for a spinning particle is not dependent on particle shape and can be reasonably approximated by the rotlet flow. Note however that the close-contact lubrication flows are dependent on the particle shape and are very difficult to calculate for gear-like spinners. 

\section{The Force Coupling Method}
The multi-body hydrodynamic interactions between the rotors are computed by the force-coupling method (FCM) \cite{Yeo10a}.
The force-coupling method is a multiscale computational model for particle suspensions in a viscous liquid, in which long-range multi-body interactions are calculated by solving the Stokes equations with a truncated, regularized multipole expansions and the singular lubrication interactions are accounted for by the analytical solutions for near-field particle-pair interactions. FCM has been successfully employed for the numerical simulations of various suspension flows \cite{Climent04,Yeo10b,Yeo10f}.

The equations of fluid motion with FCM are
\begin{eqnarray}
\bm{\nabla} p &=& \mu \nabla^2 \bm{u} + \sum^{N_p}_{n=1} \left\{ \bm{F}^n \Delta_M(\bm{r}^n)+(\bm{G}^n\cdot\bm{\nabla})\Delta_D(\bm{r}^n)\right\}, \label{Eqn:Stokes}\\
\bm{\nabla} \cdot \bm{u} &=& 0, \label{Eqn:Continuity}
\end{eqnarray}
in which $N_p$ is the number of particles, $p$ is pressure, $\mu$ is the fluid viscosity, $\bm{u}$ is the fluid velocity, $\bm{r}^n$ is a position vector measured from the center of a particle $\bm{r}^n = \bm{x} - \bm{Y}^n$, and $F_i$ and $G_{ij}$ are the force monopole and force dipole moments, respectively. The force monopole represents a body force on the particle. The dipole moment consists of anti-symmetric $C_{ij}$ and symmetric $S_{ij}$ parts. The couplet, $C_{ij}$, corresponds to a torque on the particle; $C_{ij} = \frac{1}{2} \epsilon_{ijk}T_k$. The stresslet $S_{ij}$ is related to the traction on the particle surface. The regularized multipole terms, $\Delta_M$ and $\Delta_D$, are defined as
\begin{eqnarray}
\Delta_M(\bm{r}) &=& \frac{1}{(2 \pi \sigma_M^2)^{3/2}} \exp{ \left( -\frac{r^2}{2 \sigma_M^2} \right)}, \\
\Delta_D(\bm{r}) &=& \frac{1}{(2 \pi \sigma_D^2)^{3/2}} \exp{ \left( -\frac{r^2}{2 \sigma_D^2} \right)},
\end{eqnarray}
in which $\sigma_M = a/\sqrt{\pi}$, $\sigma_D = a/(6\sqrt{\pi})^{1/3}$, and $a$ is the particle radius. Once the fluid velocity $\bm{u}$ is computed, the translational $\bm{V}$ and angular velocities $\bm{\Omega}$ are obtained from the weighted volume integral as
\begin{eqnarray}
V_i &=& \int u_i(\bm{x}) \Delta_M(\bm{r}) d^3 \bm{x}, \\
\Omega_i &=& \frac{1}{2} \int \epsilon_{ijk} \frac{\partial u_k}{\partial x_j}(\bm{x}) \Delta_D(\bm{r}) d^3 \bm{x}.
\end{eqnarray}

The force monopole and dipole moments for neutrally buoyant particles are
\begin{eqnarray}
F_i &=& F^P_i - F^{L}_i, \\
G_{ij} &=&  S^{FCM}_{ij} - C^{L}_{ij}.
\end{eqnarray}
Here, $\bm{F}^P$ is a potential force on the particles due to steric or excluded volume interactions, which is modeled by
\begin{equation} 
\bm{F}_{ij}^P =
\begin{cases}
  -6\pi\mu \dot{\gamma} a^2 F_{ref} \left( \frac{R_{ref}^2-|\bm{r}|^2}{R_{ref}^2-4a^2} \right)^6 \frac{\bm{r}}{|\bm{r}|} & \text{if $|\bm{r}| < R_{ref}$} \\
  0 & \text{otherwise},
\end{cases}
\end{equation}
in which $\bm{F}^P_{ij}$ is the force on particle $j$ by particle $i$, $\bm{r} = \bm{Y}^i - \bm{Y}^j$, $F_{ref}$ is a constant, and $R_{ref}$ is a cut-off distance. In the present study, $\bm{F}^P$ is activated if the shortest distance between two particle surfaces ($\epsilon$) is less than $0.002a$, {\it i.e.} $R_{ref}/a = 2.001$. $F_{ref}$ is chosen to keep the minimum separation distance $\epsilon_{min} \simeq 0.001a$ ($F_{ref} = 200$).
$\bm{F}^{L}$ and $\bm{C}^{L}$ are the monopole and couplet coefficients from the lubrication interaction. $\bm{S}^{FCM}$ is the set of stresslets obtained by solving (\ref{Eqn:Stokes}) subject to the rigid-body constraint that the total rate-of-strain inside of a particle is zero.

Solving (\ref{Eqn:Stokes}, \ref{Eqn:Continuity}) and computing $\bm{V}$ and $\bm{\Omega}$ from $\bm{u}$, corresponds to the following grand mobility problem,
\begin{equation}
\begin{bmatrix}
\bm{M}_{\mathcal{FU}} \bm{\mathcal{F}}^P \\ \bm{M}_{\mathcal{F}E} \bm{\mathcal{F}}^P
\end{bmatrix}
=
\begin{bmatrix}
\mathcal{R}^{-1}+\bm{M}_{\mathcal{FU}} & -\bm{M}_{S\mathcal{U}} \\
\bm{M}_{\mathcal{F}E} & -\bm{M}_{SE}
\end{bmatrix}
\begin{bmatrix}
\bm{\mathcal{F}}^{L} \\ \bm{S}^{FCM}
\end{bmatrix},
\label{LB_Symm}
\end{equation}
in which $\bm{\mathcal{F}}$ is a $(6 N_p)$ vector for the force and torque $\bm{\mathcal{F}}^T = (\bm{F}^T, \bm{T}^T)$, and the superscript $T$ denotes transpose. The local strain rate of a particle is obtained by
\begin{equation}
E_{ij} = \int e_{ij}(\bm{x}) \Delta_D(\bm{r}) d^3 \bm{x},
\end{equation}
where $e_{ij} = 1/2(\partial_j u_i + \partial_i u_j)$. $\bm{M}_{AB}$ is a FCM mobility matrix to compute a variable $B$ from a given force $A$. The FCM mobility matrix contains far-field multibody hydrodynamic interactions. Note that (\ref{LB_Symm}) is only a symbolic notation, as the far-field interaction is computed by solving (\ref{Eqn:Stokes}), instead of constructing the complicated multibody mobility matrix explicitly.

The lubrication force and torque are related to the translational and angular velocities as $\bm{\mathcal{F}}^L = \mathcal{R}\bm{\mathcal{U}}$, where $\bm{\mathcal{U}}^T = (\bm{V}^T, \bm{\Omega}^T)$. The resistance matrix $\mathcal{R}$ is defined as
\begin{equation}
\mathcal{R} =
\begin{bmatrix}
\bm{R}_{VF} & \bm{R}_{\Omega F} \\
\bm{R}_{VT} & \bm{R}_{\Omega T}
\end{bmatrix}.
\end{equation}
Here, $\bm{R}_{AB}$ is a resistance matrix relating $A$ to $B$. The resistance matrix is constructed in a pair-wise manner from the exact two-body resistance matrix, subtracting the FCM two-body resistance matrix, to account for the lubrication forces. As FCM resolves the far-field hydrodynamic interaction almost exactly, $|\bm{R}_{AB}| \rightarrow 0$ for the center-to-center distance of a particle pair $r > 0.6a$.
The FCM stresslet $\bm{S}^{FCM}$ is related to the physical stresslet $\bm{S}$ as
\begin{equation}
\bm{S}^{FCM} = \bm{S}-\bm{R}_{\mathcal{U}S} \mathcal{U}.
\label{Stresslet_Eq}
\end{equation}
Note that, in the computation, we solve for $\bm{S}^{FCM}$, not $\bm{S}$. $\bm{S}$ for the system of particles can be calculated as a post-processing step. \cite{Yeo10a} proposed an efficient preconditioned conjugate gradient solver for (\ref{LB_Symm}), in which the number of floating point operations in one iteration is almost the same as the standard FCM.
For details, see Yeo \& Maxey \cite{Yeo10a}.


\begin{thebibliography}{99}
\bibitem{quasi2D}
Note that even though each rotor particle moves only within the monolayer, mimicking a 2D system, the particle motion is affected by the  full 3D hydrodynamic interactions though the embedding fluid, i.e., the flows in the fluid space above and below the monolayer.
\bibitem{rcp}
For a monolayer random close packing is set by rotors cross-sectional area fraction $\phi_A\sim 0.84$. $\phi=\phi_A 2/3$
\end{thebibliography}
\end{document}